\documentclass[10pt, doublecolumn]{IEEEtran}
\usepackage{epsfig,latexsym}
\usepackage{float}
\usepackage{indentfirst}
\usepackage{amsmath}
\usepackage{bm}
\usepackage{amssymb}
\usepackage{times}
\usepackage{enumitem}

\usepackage{algorithm}
\usepackage[noend]{algpseudocode}
\usepackage{subfigure}
\usepackage{psfrag}
\usepackage{hyperref}
\usepackage{cite}
\usepackage{lastpage}
\usepackage{fancyhdr}
\usepackage{color} 
 \usepackage{amsthm}
\usepackage{bigints}
\usepackage{array}
\usepackage{booktabs}
\newcounter{problem}
\newcounter{save@equation}
\newcounter{save@problem}
\makeatletter
\newenvironment{problem}
{\setcounter{problem}{\value{save@problem}}%
  \setcounter{save@equation}{\value{equation}}%
  \let\c@equation\c@problem
  \subequations
}
{\endsubequations
  \setcounter{save@problem}{\value{equation}}%
  \setcounter{equation}{\value{save@equation}}%
}

\begin{document}
\title{  \vspace{-1em}{ Toward a Quiet Wireless World: Multi-Cell Pinching-Antenna Transmission    }}

\author{ Zhiguo Ding, \IEEEmembership{Fellow, IEEE}  \thanks{ 
  
\vspace{-1em}

Z. Ding is with the School of Electrical and Electronic Engineering, Nanyang Technological University, Singapore.  
 

  }\vspace{-3em}}
 \maketitle

\begin{abstract}
Conventional-antenna-based multi-cell interference management can lead to excessive power consumption. For example, in order to serve those users which are close to the cell edge, base stations often must transmit at very high power levels to overcome severe large-scale path-loss, i.e., the base stations have to ``shout" at the users to realize the users' target quality of service (QoS). This letter focuses on the application of pinching antennas to multi-cell interference management and demonstrates that the use of multi-cell pinching-antenna transmission leads to a quiet wireless world. In particular, each transceiver pair can be positioned in close proximity, and hence the users' QoS requirements can be met with only low transmit power, i.e., via ``whispering" rather than high-power transmission. 
\end{abstract}\vspace{-0.5em}

\vspace{-1em} 

\section{Introduction}
Conventional cellular networks are essentially interference-limited systems, which makes interference management a highly power-consuming task \cite{Rappaport}. For example, to serve those users which are close to the cell edge, base stations often must transmit at very high power levels to overcome severe large-scale path-loss, i.e., the base stations have to ``shout" at the users to realize the users' target quality of service (QoS) requirements. To mitigate such large-scale path-loss, various novel technologies, such as cloud radio access networks (C-RAN), cell-free massive multiple-input multiple-output (MIMO), coordinated multipoint (CoMP), etc, have been developed \cite{CMCC,5706317,you6g}. These technologies typically rely on data sharing and extensive coordination among the base stations, which leads to substantial backhaul/fronthaul burdens and stringent time/phase synchronization challenges.

This letter focuses on the impact of the recently developed technology, namely pinching antennas, on the multi-cell interference management \cite{pinching_antenna2}. The key idea of pinching-antenna systems is to utilize the feature of dielectric waveguides, that radio-frequency signals can be leaked at any point along a waveguide, which means that a base station can be activated at a location close to its serving user \cite{mypa}. While pinching antennas have been extensively studied in the single-cell context, their impact on the multi-cell system has not yet been fully investigated, although the initial two studies in \cite{11195162,11315149} demonstrated the great potential of multi-cell pinching-antenna systems from a stochastic geometric perspective.  

The main contribution of this letter is to design multi-cell pinching-antenna transmission from an optimization perspective. In particular, a practical system model is established for multi-cell pinching-antenna transmission, where data sharing among base stations is not required and there is no need for time/phase synchronization among the base stations.  Then, an overall transmit power minimization problem is formulated for the addressed mult-cell pinching-antenna system. Because of the non-convex nature of the formulated optimization problem, cross-entropy optimization is applied by using the prior knowledge of the optimal locations of the pinching antennas. For the two-cell special case, the infeasibility issue, which is inherent to multi-cell systems, is investigated, and a suboptimal solution for the antenna placement optimization is also proposed. The simulation results are also presented to reveal that interference management can be realized without stringent coordination among the cells, and the use of multi-cell pinching-antenna transmission leads to a quiet wireless world. In particular, each transceiver pair can be positioned in close proximity, and hence the users' QoS requirements can be met with only low transmit power, i.e., via ``whispering" rather than high-power transmission. 
\vspace{-1em}
\section{System Model}\label{section model}
This letter focuses on a multi-cell downlink communication scenario assisted by pinching antennas. In particular, consider a rectangular-shaped service area, with its two sides denoted by $D_{\rm L}$ and $D_{\rm W}$, respectively. The service area is divided into $N_{\rm row}$ rows and $N_{\rm col}$ columns, which means that there are $M=N_{\rm row}N_{\rm col}$ equal-sized rectangular-shaped cells\footnote{Note that the circular or hexagonal shapes have been used in conventional cellular systems with fixed-location base stations. Because of the flexible locations of pinching antennas, rectangular-shaped cells are used in this paper.  }.

For the conventional-antenna benchmarking scheme, the base stations are placed at the centers of the cells, i.e., the location of the base station in the cell on the $i$-th row and the $j$-th column is given by $ {\boldsymbol \psi}_{i,j}^{\rm Conv}=
\left(
-\frac{D_{\rm L}}{2}+\left(j-\frac12\right)\frac{D_{\rm L}}{N_{\rm col}},
-\frac{D_{\rm W}}{2}+\left(i-\frac12\right)\frac{D_{\rm W}}{N_{\rm row}},d
\right)$, where $d$ denotes the height of the antenna. For the pinching-antenna assisted multi-cell system, assume that a dielectric waveguide with a length of $\frac{D_{\rm L}}{N_{\rm col}}$ is placed at the center of each cell, with a single pinching antenna activated on each waveguide.
 
In order to simplify the notations, the cell on the $i$-th row and the $j$-th column is denoted by ${\rm Cell}_m$, $m=(i-1)N_{\rm row}+j$, $1\leq m \leq M$. The pinching antenna and the user in ${\rm Cell}_m$ are denoted by ${\rm PA}_m$ and ${\rm U}_m$, respectively. The coordniates of the conventional antenna in ${\rm Cell}_m$, ${\rm PA}_m$ and ${\rm U}_m$ are denoted by $ {\boldsymbol \psi}_{m}^{\rm Conv}=
\left(x_{m}^{\rm Conv}, y_{m}^{\rm Conv},d
\right)$, ${\boldsymbol \psi}_{m}^{\rm Pin}=\left(x_{m}^{\rm Pin}, y_{m}^{\rm Conv},d\right)$ and ${\boldsymbol \psi}_{m}=\left(x_{m}, y_{m},0\right)$, respectively, where $x_{m}^{\rm Conv}$ and $y_{m}^{\rm Conv}$ can be obtained from ${\boldsymbol \psi}_{i,j}^{\rm Conv}$ straightforwardly.

Therefore, for the considered multi-cell downlink scenario, the signal received by  ${\rm U} _m$  is given by
 \begin{align}\label{models}
  y_m=&    {h}_{mm}   s_m +   \sum_{i=1,i\neq m}^{M}{h}_{im}  s_i +w_m  ,
  \end{align}
  where $s_m$ denotes the signal sent by ${\rm PA}_m$, $w_m$ denotes the additive white Gaussian noise with its variance denoted by $P_{\rm N}$, 
    $h_{im}$ denotes the channel between ${\rm PA} _i^{\rm Pin}$ and ${\rm U}_m$ as follows: 
 \begin{align}
 h_{im}=\frac{\sqrt{\eta} e^{-2\pi j \left(\frac{  1}{\lambda}\left| {\boldsymbol \psi}_m  - {\boldsymbol \psi}_i^{\rm Pin}\right|
  +\frac{1}{\lambda_g}\left| {\boldsymbol \psi}_{0i}^{\rm Pin}  - {\boldsymbol \psi}_i^{\rm Pin}\right|
  \right)}}{  \left| {\boldsymbol \psi} _m - {\boldsymbol \psi}_i^{\rm Pin}\right|} ,
  \end{align}
  $\eta = \frac{c^2}{16\pi^2 f_c^2 }$, $c$ is the speed of light, $f_c$ is the carrier frequency, $\lambda=\frac{c}{f_c}$, ${\boldsymbol \psi}_{0m}^{\rm Pin} $ is the location of the feed point of the $m$-th waveguide,  and $\lambda_g$ denotes the waveguide wavelength \cite{mypa}.

  Therefore, ${\rm U}_m$'s achievable downlink data rate in the considered multi-cell scenario is given by
  \begin{align}\label{eq4}
  R_m  =& \log_2\left(
1+ \frac{ P_m\left|h_{mm} \right|^2}
{    \sum_{i=1,i\neq m}^{M}P_i\left|
h_{im}   
\right|^2+P_{\rm N}}
\right)\\\nonumber
=&
\log_2\left(
1+ \frac{ P_m \frac{\eta}{ \left| {\boldsymbol \psi} _m - {\boldsymbol \psi}_m^{\rm Pin}\right|^2}}
{    \sum_{i=1,i\neq m}^{M}P_i\frac{\eta}{ \left| {\boldsymbol \psi} _im- {\boldsymbol \psi}_i^{\rm Pin}\right|^2}+P_{\rm N}}
\right),
\end{align} 
where $P_m$ denotes ${\rm PA}_m$'s transmit power for ${\rm U}_m$'s signal.  
 
 \section{Power and Antenna Location Optimization} 
 The aim of this letter is to minimize the overall transmit power consumption for the considered multi-cell pinching-antenna system, where the optimization variables include the transmit powers and the locations of the pinching antennas. In particular, the considered optimization problem can be formulated as follows:
 \begin{problem}\label{pb:1} 
  \begin{alignat}{2}
\underset{ P_m, x_m^{\rm Pin} }{\rm{min}}  &\quad   \sum_{m=1}^{M}P_m
\\ s.t. &\quad  \label{1tst:1}  R_m\geq R_t, \quad 1\leq m \leq M
\\ &\quad x_{m,{\rm start}}^{\rm Pin}\leq x_m^{\rm Pin}\leq x_{m,{\rm end}}^{\rm Pin}   , \quad 1\leq m \leq M, 
  \end{alignat}
\end{problem}
where we assume that the users have the same target data rate, denoted by  $R_t$, $x_{m,{\rm start}}^{\rm Pin}$ and $x_{m,{\rm end}}^{\rm Pin}$ denote the boundaries of the location of ${\rm PA}_m$, respectively.
  
Because of the non-convex nature of problem \eqref{pb:1}, in the following, a closed-form solution of power allocation is first obtained by treating the antenna locations as constant. Then, by using the closed-form power allocation solution,  problem \eqref{pb:1} can be simplified to an antenna location optimization problem, where cross-entropy optimization will be applied \cite{crossentropy}.  

In particular, with fixed $x_m^{\rm Pin}$, problem \eqref{pb:1} can be expressed as follows:  
        \begin{problem}\label{pb:2} 
  \begin{alignat}{2}
\underset{ P_m }{\rm{min}}  &\quad   \sum_{m=1}^{M}P_m
\\ s.t. &\quad  \nonumber   P_m\left|h_{mm} \right|^2 - \epsilon \left(\sum^{M}_{i=1,i\neq m}P_i\left|
h_{im}   
\right|^2\right)-\epsilon P_{\rm N}
  \geq 0,    \\&\quad \quad 1\leq m\leq M, \label{1tst:2} 
  \end{alignat}
\end{problem}  
where $\epsilon=2^{R_t}-1$. Problem \eqref{pb:2} is a linear programming problem, and hence its optimal solution can be obtained by applying the Karush–Kuhn–Tucker (KKT) conditions \cite{Boyd}. In particular, the Lagrange of problem \eqref{pb:2} is given by 
\begin{align}
L =&  \sum_{m=1}^{M}P_m-\sum^{M}_{m=1}\lambda_m\\\nonumber &\times\left(P_m\left|h_{mm} \right|^2 - \epsilon \left(\sum^{M}_{i=1,i\neq m}P_i\left|
h_{im}   
\right|^2\right)-\epsilon P_{\rm N}\right),
\end{align}
where $\lambda_m$ denotes the Lagrangian multiplier, and the partial derivative of the Lagrange with respect to $P_k$ is given by
\begin{align}
\frac{\partial L}{\partial P_k} = 1-  \lambda_k  \left|h_{kk} \right|^2 +\sum^{M}_{m=1,m\neq k}\lambda_m \epsilon \left|
h_{km}   
\right|^2  =0.
\end{align}
By applying the stationarity condition, it is straightforward to show that $\lambda_m\neq 0$, $m=\{1,\cdots, M\}$.  Therefore, by applying the complementary slackness condition, the constraints in \eqref{1tst:2} must hold with equality, i.e., the optimal solution can be obtained by solving the following equations: 
\begin{align}
 P_m\left|h_{mm} \right|^2 - \epsilon \left(\sum^{M}_{i=1,i\neq m}P_i\left|
h_{im}   
\right|^2\right)-\epsilon P_{\rm N}=0,  
\end{align} 
which leads to the following linear equations: 
\begin{align}\label{linear equation}
\left(\mathbf{I}_M-\epsilon \mathbf{G}\right)\mathbf{p}=\mathbf{a} ,
\end{align}
where $\mathbf{p} =  \begin{bmatrix}P_1&\cdots &P_M
\end{bmatrix}^T$, $\mathbf{a} = \epsilon P_{\rm N}\begin{bmatrix}
\frac{1}{\left|
h_{11}   
\right|^2}&\cdots &\frac{1}{\left|
h_{MM}   
\right|^2}
\end{bmatrix}^T$, and 
\begin{align}
\mathbf{G}=\begin{bmatrix}
0 & \frac{ \left|
h_{21}   
\right|^2}{ \left|
h_{11}   
\right|^2} &\cdots &\frac{\left|
h_{(M-1)1}   
\right|^2}{ \left|
h_{11}   
\right|^2}&\frac{\left|
h_{M1}   
\right|^2}{ \left|
h_{11}   
\right|^2}\\
\frac{   \left| h_{12}
\right|^2}{ \left|
h_{22}   
\right|^2} &  0 &\cdots & \frac{\left|
h_{(M-1)2}   
\right|^2}{ \left|
h_{22}   
\right|^2}& \frac{\left|
h_{M2}   
\right|^2}{ \left|
h_{22}   
\right|^2}\\\vdots &\ddots&\ddots&\ddots&\vdots\\
\frac{ \left|
h_{1M}   
\right|^2}{ \left|
h_{MM}   
\right|^2} & \frac{ \left|
h_{2M}   
\right|^2}{ \left|
h_{MM}   
\right|^2} &\cdots & \frac{\left|
h_{(M-1)M}   
\right|^2}{ \left|
h_{MM}   
\right|^2} & 0
\end{bmatrix}.
\end{align}

Assume that $\mathbf{I}_M-\epsilon \mathbf{G} $ is inverable and each element of $\left(\mathbf{I}_M-\epsilon \mathbf{G}\right)^{-1}\mathbf{a}$ is positive, problem \eqref{pb:1} can be equvalently expressed as follows: 
 \begin{problem}\label{pb:3} 
  \begin{alignat}{2}\label{ob:1}
\underset{   x_m^{\rm Pin} }{\rm{min}}  &\quad    {\rm sum}\left\{  \left(\mathbf{I}_M-\epsilon \mathbf{G}\right)^{-1}\mathbf{a}
\right\} 
\\ s.t. &\quad  \label{1tst:3}  x_{m,{\rm start}}^{\rm Pin}\leq x_m^{\rm Pin}\leq x_{m,{\rm end}}^{\rm Pin}   , \quad 1\leq m \leq M, 
  \end{alignat}
\end{problem}
where   ${\rm sum}$ denotes the sum operation for a vector.  
 
Compared to problem \eqref{pb:1}, problem \eqref{pb:3} is simpler since the problem is a function of the antenna locations only. However, the analytical optimal solution of problem \eqref{pb:3} is still challenging to obtain, due to its non-convex nature, which motivates the cross-entropy algorithm and the two-cell special case study presented in the following two subsections. 
\vspace{-1em}
\subsection{Cross-Entropy Optimization}
 The motivation to use cross-entropy optimization is two-fold. Firstly, the use of cross-entropy optimization does not require the first-order derivative of the objective function with respect to the optimization variables, which is ideal to problem \eqref{pb:3}, given the complex expression shown in \eqref{ob:1}. Secondly, there is prior knowledge available about the optimal location of ${\rm PA}_m$, e.g., it is likely that ${\rm PA}_m$ is around ${\rm U}_m$. This knowledge can be effectively utilized by cross-entropy optimization.  

The key idea of cross-entropy optimization is to learn/optimize a distribution whose samples are close-to-optimal solutions of the optimization problem \cite{crossentropy}. Denote this ideal distribution by $p^*\left(\mathbf{x}^{\rm Pin}\right)$, where $\mathbf{x}^{\rm Pin}$ is an $M\times1$ vector collecting all the antenna locations. The aim of cross-entropy optimization is to refine a parametric distribution, denoted by $p\left(\mathbf{x}^{\rm Pin};{\boldsymbol \theta}\right)$, where ${\boldsymbol \theta}$ are the parameters to be optimized. The criterion for cross-entropy optimization to refine $p\left(\mathbf{x}^{\rm Pin};{\boldsymbol \theta}\right)$ is the Kullback–Leibler (KL) divergence which measures the difference between  $p^*\left(\mathbf{x}^{\rm Pin}\right)$ and $p\left(\mathbf{x}^{\rm Pin};{\boldsymbol \theta}\right)$, i.e., cross-entropy optimization is to minimize the following:
 \begin{problem}\label{pb:4} 
  \begin{alignat}{2}\label{ob:14}
\underset{{\boldsymbol \theta}}{\min} &\quad D_{\rm KL}(p^*  \| p )
\triangleq  \int p^*\left(\mathbf{x}^{\rm Pin}\right) \log\frac{p^*\left(\mathbf{x}^{\rm Pin}\right)}{p\left(\mathbf{x}^{\rm Pin};{\boldsymbol \theta}\right)}d{\boldsymbol \theta}.
  \end{alignat}
\end{problem}
 With some straightforward algebraic manipulations, problem \eqref{pb:4} is equivalent to the following entropy-minimization problem: 
  \begin{problem}\label{pb:5} 
  \begin{alignat}{2}\label{ob:15}
\underset{{\boldsymbol \theta}}{\min} \quad  
H(p^*  \| p )
= -\int p^*\left(\mathbf{x}^{\rm Pin}\right)  \log p\left(\mathbf{x}^{\rm Pin};{\boldsymbol \theta}\right)d{\boldsymbol \theta},
  \end{alignat}
\end{problem}
which can also be written as follows:
  \begin{problem}\label{pb:6} 
  \begin{alignat}{2}\label{ob:6}
\underset{{\boldsymbol \theta}}{\max} \quad  
\mathcal{E}_{p^*}\left\{ \log p\left(\mathbf{x}^{\rm Pin};{\boldsymbol \theta}\right) \right\}.
  \end{alignat}
\end{problem}
If $p^*\left(\mathbf{x}^{\rm Pin}\right)$ is known, a maximum likelihood estimator (MLE) can be applied to the samples drawn from $p^*\left(\mathbf{x}^{\rm Pin}\right)$ and then solve problem \eqref{pb:6}. Since   $p^*\left(\mathbf{x}^{\rm Pin}\right)$ is not known, the good sample which fits the requirement the best, termed elite samples, are used as the samples for the estimation.

 \begin{algorithm}[t]
\caption{Cross-Entropy Optimization Algorithm}

 \begin{algorithmic}[1]
 
\State Initilization: ${\boldsymbol \mu}_{(1)}=\bar{\mathbf{x}}$, ${\boldsymbol \sigma}_{(1)}=\frac{D_{\rm L}}{2N_{\rm Col}}\mathbf{1}$  
\For  {$n = 0,1,\dots,N_{\max}$}  
\State $n=n+1$
\State Draw $N_{\rm CE}$ i.i.d. samples from $\mathcal{N}({\boldsymbol \mu}_{(n-1)}, {\boldsymbol \sigma}_{(n-1)}^2)$

\State Carry out boundary projection

\State Compute the overall transmit power by using \eqref{linear equation}

\State Construct a set $\mathcal{S}$ containing $N_{\rm best}$ elite samples  

yielding the smallest transmit power

\State Use the elite samples to generate new estimates for 

the mean and the variance, ${\boldsymbol \mu}_{\rm New}$ and ${\boldsymbol \sigma}_{\rm New}^2$

\State Update the mean with the smoothing step

${\boldsymbol \mu}_{(n)}= \alpha {\boldsymbol \mu}_{(n-1)}+(1-\alpha){\boldsymbol \mu}_{\rm New}$

\State Update the variance with smoothing and additive noise

${\boldsymbol \sigma}_{(n)}^2= \alpha {\boldsymbol \sigma}_{(n-1)}^2+(1-\alpha){\boldsymbol \sigma}_{\rm New}^2+ \delta$

 \EndFor
 
\State \textbf{end}
 \State Select the sample that yields the smallest transmit power as the antenna locations.

 \end{algorithmic}\label{algorithm1}
\end{algorithm}

Algorithm \ref{algorithm1} outlines the key steps for the implementation of the proposed cross-entropy optimization algorithm. In particular, the means of the antenna locations, $x^{\rm Pin}_m$, are initialized by using $\bar{\mathbf{x}}$ whose $m$-th element is simply $x_m$. The initialization for the variances is based on the width of each waveguide. The proposed algorithm carries out $N_{\rm max}$ iterations. During the $n$-th iteration,  $N_{\rm CE}$ independent and identically distributed (i.i.d.) samples are dawn from the Gaussian distribution by using the mean and the variance from the previous iteration, i.e.,  $\mathcal{N}\left({\boldsymbol \mu}_{(n-1)}, {\boldsymbol \sigma}_{(n-1)}^2\right)$. Denote these samples by $\tilde{\mathbf{x}}_k$, $1\leq k \leq N_{\rm CE}$. Since the generated samples might fall out of the feasible region, a boundary project is carried out to force that each antenna location is within the corresponding feasible region, i.e., $x_{m,{\rm start}}^{\rm Pin}\leq x_m^{\rm Pin}\leq x_{m,{\rm end}}^{\rm Pin} $. 

Then,  $N_{\rm best}$ elite samples which yield the smallest transmit power are selected to be included in the elite subset $\mathcal{S}$. These elite samples can be used to generate new estimates for 
the mean and the variance,
denoted by ${\boldsymbol \mu}_{\rm New}$ and ${\boldsymbol \sigma}_{\rm New}^2$, respectively, i.e.,
\begin{align}
{\boldsymbol \mu}_{\rm New}=\frac{\sum_{k\in\mathcal S}\tilde{\mathbf{x}}_k}{N_{\rm CE}},\qquad
{\boldsymbol \sigma}_{\rm New}^2=\frac{\sum_{k\in\mathcal S}(\tilde{\mathbf{x}}_k-{\boldsymbol \mu}_{\rm New})^2}{N_{\rm CE}}.
\end{align}

At the end of the $n$-th iteration, both the mean and the variance will be updated by using the newly obtained estimates, ${\boldsymbol \mu}_{\rm New}$ and ${\boldsymbol \sigma}_{\rm New}^2$, as well as the smoothing step, as shown in Algorithm \ref{algorithm1}, where $\alpha$ denotes the smoothing parameter, and $\delta$ denotes the additive smoothing noise. 
 \vspace{-1em}

\subsection{Two-Cell Speical Case}
The aim of this subsection is to focus on the two-cell special case, where additional insight into the performance of multi-cell pinching-antenna systems can be obtained. 

By using the solution shown in \eqref{linear equation}, for fixed antenna locations, the optimal solution of the transmit power allocation for the two-cell special case is given by
  \begin{align} 
P_1^*=&
\frac{\epsilon P_{\rm N}\left(|h_{22}|^2+\epsilon |h_{21}|^2\right)}
{|h_{11}|^2|h_{22}|^2-\epsilon^2 |h_{21}|^2|h_{12}|^2},\\
P_2^*=&
\frac{\epsilon P_{\rm N}\left(|h_{11}|^2+\epsilon |h_{12}|^2\right)}
{|h_{11}|^2|h_{22}|^2-\epsilon^2 |h_{21}|^2|h_{12}|^2},
\end{align} 
if $|h_{11}|^2|h_{22}|^2-\epsilon^2 |h_{21}|^2|h_{12}|^2>0$, otherwise it is infeasible to achieve the target data rate $R_t$. 
Therefore, the overall transmit power required by the pinching-antenna system is given by
  \begin{align} \label{two-user1}
P_1^*+P_2^* =
\frac{\epsilon P_{\rm N}\left(|h_{11}|^2+|h_{22}|^2+\epsilon\left(|h_{21}|^2+|h_{12}|^2\right)\right)}
{|h_{11}|^2|h_{22}|^2-\epsilon^2 |h_{21}|^2|h_{12}|^2},
\end{align} 
which can be further written as an explicit function of the antenna locations as shown in \eqref{p1p2} on the top of the next page, where $\xi=y_1^2+y_2^2+2d^2$, and $f\left({x}_1^{\rm Pin},{x}_2^{\rm Pin}\right)=\frac{\left(\left( {x}_1^{\rm Pin}-x_1\right)^2+y_1^2+d^2\right)\left(\left( {x}_2^{\rm Pin}-x_2\right)^2+y_2^2+d^2\right)}{\left(\left( {x}_2^{\rm Pin}-x_1\right)^2+y_1^2+d^2\right)\left(\left( {x}_1^{\rm Pin}-x_2\right)^2+y_2^2+d^2\right)}$. 

\begin{figure*}\vspace{-2em}
  \begin{align} \label{p1p2}
P_1^*+P_2^* =
\frac{\eta^{-1}\epsilon P_{\rm N}\left(\left( {x}_1^{\rm Pin}-x_1\right)^2+\left( {x}_2^{\rm Pin}-x_2\right)^2+\xi+\epsilon f\left({x}_1^{\rm Pin},{x}_2^{\rm Pin}\right) \left(\left( {x}_1^{\rm Pin}-x_2\right)^2+\left( {x}_2^{\rm Pin}-x_1\right)^2+\xi\right)\right)}
{1-\epsilon^2 f\left({x}_1^{\rm Pin},{x}_2^{\rm Pin}\right)}.
\end{align} \vspace{-2em}
\end{figure*}

As can be seen from \eqref{p1p2}, the overall transmit power is still a complex non-convex function of the antenna locations, ${x}_1^{\rm Pin}$ and ${x}_2^{\rm Pin}$. Therefore, a closed-form analytical expression is challenging to obtain even for the two-cell special case; however, a few insightful understandings about multi-cell pinching-antenna transmission can be obtained, as follows.

\subsubsection{Feasibility}  The condition to ensure that the target data rate is achievable is given by $ |h_{11}|^2|h_{22}|^2-\epsilon^2 |h_{21}|^2|h_{12}|^2\geq 0$, which imposes an upper bound on the target data rate as follows:
 {\small \begin{align} \label{loerdd}
  R_t\leq \log\left(1+\sqrt{ \frac{ |h_{11}|^2|h_{22}|^2}{|h_{21}|^2|h_{12}|^2}}\right)= \log\left(1+\frac{1}{\sqrt{  f\left({x}_1^{\rm Pin},{x}_2^{\rm Pin}\right)}}\right). 
\end{align} }

For conventional-antenna systems, the channel gains are decided by the wireless propagation environment, and hence are non-configurable parameters. As a result, for many users, particularly cell-edge users, infeasibility occurs frequently. However, by using pinching antennas, the infeasibility situation can be mitigated by changing the locations of the pinching antennas. For example, by placing ${\rm PA}_1$ close to ${\rm U}_1$ but far away from ${\rm U}_2$, the ratio $\frac{ |h_{11}|^2 }{|h_{21}|^2}$ can be increased, which is helpful to reduce the occurrence of infeasibility.  

\subsubsection{A suboptimal solution with a closed form expression}\label{subsection suboptimal}
Recall that by using \eqref{p1p2}, problem \eqref{pb:1} can be rewritten as shown in \eqref{pb:7} on the top of the next page. 
\begin{figure*}\vspace{-2em}
 \begin{problem}\label{pb:7} 
  \begin{alignat}{2}\label{ob:17}
\underset{{x}_m^{\rm Pin}}{\min} &\quad  
\frac{ \left( {x}_1^{\rm Pin}-x_1\right)^2+\left( {x}_2^{\rm Pin}-x_2\right)^2+\xi+\epsilon f\left({x}_1^{\rm Pin},{x}_2^{\rm Pin}\right) \left(\left( {x}_1^{\rm Pin}-x_2\right)^2+\left( {x}_2^{\rm Pin}-x_1\right)^2+\xi\right)}
{1-\epsilon^2 f\left({x}_1^{\rm Pin},{x}_2^{\rm Pin}\right)}
\\ s.t. &\quad  \label{1tst:3}  -\frac{D_{\rm L}}{2}\leq x_1^{\rm Pin}\leq 0  ,\quad 0\leq x_2^{\rm Pin}\leq \frac{D_{\rm L}}{2}.
  \end{alignat}
\end{problem}
  \end{figure*}

We note that the term $f\left({x}_1^{\rm Pin},{x}_2^{\rm Pin}\right)$ plays an important role for the optimization problem shown in \eqref{pb:7}. For example, one can view that the term, $\left( {x}_1^{\rm Pin}-x_1\right)^2+\left( {x}_2^{\rm Pin}-x_2\right)^2+\xi$, in the objective function of problem \eqref{pb:7} as the signal part to be boosted, and  $\left( {x}_1^{\rm Pin}-x_2\right)^2+\left( {x}_2^{\rm Pin}-x_1\right)^2+\xi$ as the interference part to be suppressed. Therefore, decreasing the value of $f\left({x}_1^{\rm Pin},{x}_2^{\rm Pin}\right)$ is preferable since the ratio between the signal and interference terms is increasing. Furthermore, decreasing $f\left({x}_1^{\rm Pin},{x}_2^{\rm Pin}\right)$ also leads to a direct consequence of an increased upper bound on the target data rate, as shown in \eqref{loerdd}. These observations lead to the following optimization problem:         
 \begin{problem}\label{pb:71} 
  \begin{alignat}{2}\label{ob:171}
\underset{{x}_m^{\rm Pin}}{\min} &\quad  f\left({x}_1^{\rm Pin},{x}_2^{\rm Pin}\right)
\\ s.t. &\quad  \label{1tst:3}  -\frac{D_{\rm L}}{2}\leq x_1^{\rm Pin}\leq 0  ,\quad 0\leq x_2^{\rm Pin}\leq \frac{D_{\rm L}}{2}.
  \end{alignat}
\end{problem}
We note that the two cells' parameters are decoupled in the objective function of problem \eqref{pb:71}, which means problem \eqref{pb:71} can be decomposed as the following two optimization sub-problems: 
        \begin{problem}\label{pb:8} 
  \begin{alignat}{2}
\underset{   x_1^{\rm Pin} }{\rm{max}}  &\quad  g_1\left(x_1^{\rm Pin}\right)\triangleq \frac{  \left( x_1^{\rm Pin}-x_2  \right)^2 +y_2^2+d^2  }{  \left( x_1^{\rm Pin}-x_1  \right)^2 +y_1^2+d^2   } 
\\ s.t. &\quad  \label{1tst:1}   -\frac{D_{\rm L}}{2}\leq x_1^{\rm Pin}\leq x_1  ,  
  \end{alignat}
\end{problem}  
and
        \begin{problem}\label{pb:9} 
  \begin{alignat}{2}
\underset{   x_2^{\rm Pin} }{\rm{max}}  &\quad  g_2\left(x_1^{\rm Pin}\right)\triangleq \frac{  \left( x_2^{\rm Pin}-x_1  \right)^2 +y_1^2+d^2  }{  \left( x_2^{\rm Pin}-x_2  \right)^2 +y_2^2+d^2   } 
\\ s.t. &\quad  \label{1tst:1}   x_2\leq x_2^{\rm Pin}\leq \frac{D_{\rm L}}{2},
  \end{alignat}
\end{problem}  
where $x_1$ is used as the upper bound on $x_1^{\rm Pin}$ and $x_2$ is used as the lower bound on $x_2^{\rm Pin}$, as explained in the following. Consider a choice of $x_1^{\rm Pin}=x_1+t$, where $t$ has a small positive value. It is straightforward to show that the choice of $x_1^{\rm Pin}=x_1-t$ outperforms $x_1^{\rm Pin}=x_1+t$ since $x_1^{\rm Pin}=x_1-t$ reduces the interference channel gain $|h_{12}|^2$ but maintains the desirable channel gain $|h_{11}|^2$.

The two optimization problems in \eqref{pb:8} and \eqref{pb:9} can be solved in the same manner. So without loss of generality, problem \eqref{pb:8} is focused on. Although the concavity of the objective function of problem \eqref{pb:8} cannot be established, the optimal solution of problem \eqref{pb:8} can be one of the following. One possible solution is from the boundary points, i.e., $-\frac{D_{\rm L}}{2}$ and $x_1$, and the other is from the stationary points of the objective function $ g_1\left(x_1^{\rm Pin}\right)$. Note that the first-order derivative of the objective function is given by
\begin{align} 
\frac{d g\left(x_1^{\rm Pin}\right)}{d x_1^{\rm Pin}}
&=
\left( (x_1^{\rm Pin}-x_1)^2+y_1^2+d^2 \right)^{-2}\\\nonumber &\times\left(
2(x_1^{\rm Pin}-x_2) 
\Big[(x_1^{\rm Pin}-x_1)^2+y_1^2+d^2\Big]
-
\right.\\\nonumber &\left. 2(x_1^{\rm Pin}-x_1)
\Big[(x_1^{\rm Pin}-x_2)^2+y_2^2+d^2\Big]\right).
\end{align}
Therefore, the stationary points could be in one of the following two forms:
\begin{align}\label{stationary}
x_{1}^{\rm Pin}
=
\frac{-\gamma_2 \pm \sqrt{\gamma_2^2 - 4\gamma_1\gamma_3}}
{2\gamma_1},
\end{align}
where $\gamma_1 = x_2-x_1$, $\gamma_2 = -(x_2-x_1)(x_1+x_2) + (y_1^2-y_2^2)$, and $\gamma_3 = (x_2-x_1)x_1x_2
-(y_1^2+d^2)x_2
+(y_2^2+d^2)x_1$. Here, it is assumed that $\gamma_2^2 \geq 4\gamma_1\gamma_3$, otherwise, there is no stationary point. Therefore, by comparing the values of the objective function with the two boundary points and the stationary points shown in \eqref{stationary}, the optimal solution of problem \eqref{pb:8} can be obtained without carrying out any iterations. 

It is interesting to note that the stationary points of $ g_1\left(x_1^{\rm Pin}\right)$ are the same as those of $ g_2\left(x_1^{\rm Pin}\right)$, which means that the optimal solution of problem \eqref{pb:9} can be obtained straighforwardly after problem \eqref{pb:8} is solved.  
\section{Numerical Studies}
In this section, the proposed multi-cell pinching-antenna scheme is evaluated via computer simulations, where $D_{\rm W}=20$ m,  $D_{\rm L}= 4D_{\rm W}$, the noise power is $-70$ dBm, $N_{\rm CE} = 1000$, $N_{\rm max} = 10$, $N_{\rm best} = 10$, $\alpha=0.7$, and $ \delta = 0.1$. 
   \begin{figure}[!] \vspace{-2em}
\begin{center}
\subfigure[ Use $P_{\rm max}=50$ dBm for the infeasibility situations ]{\label{fig1a}\includegraphics[width=0.3\textwidth]{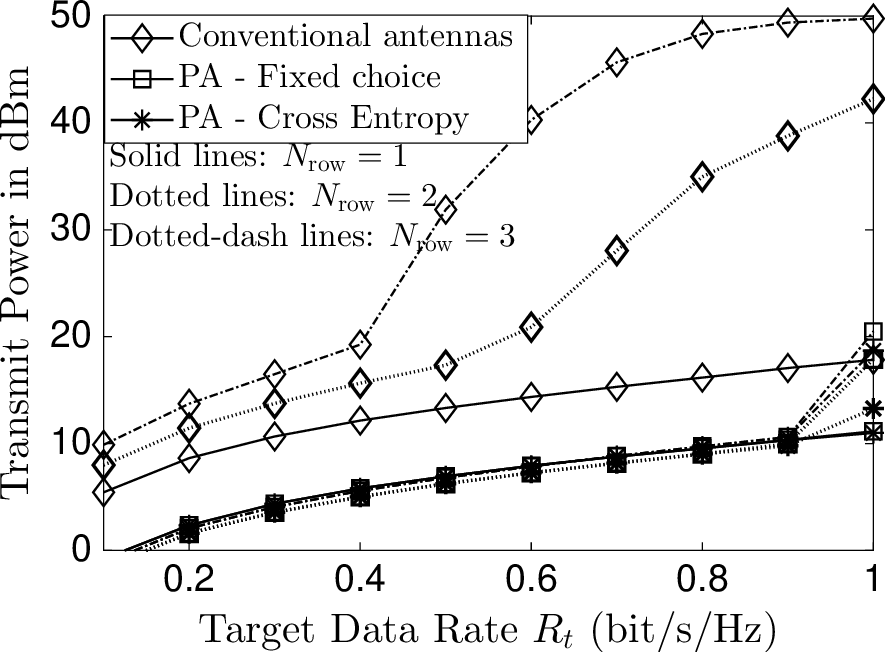}} 
\subfigure[Simply discard the infeasibility situations ]{\label{fig1b}\includegraphics[width=0.3\textwidth]{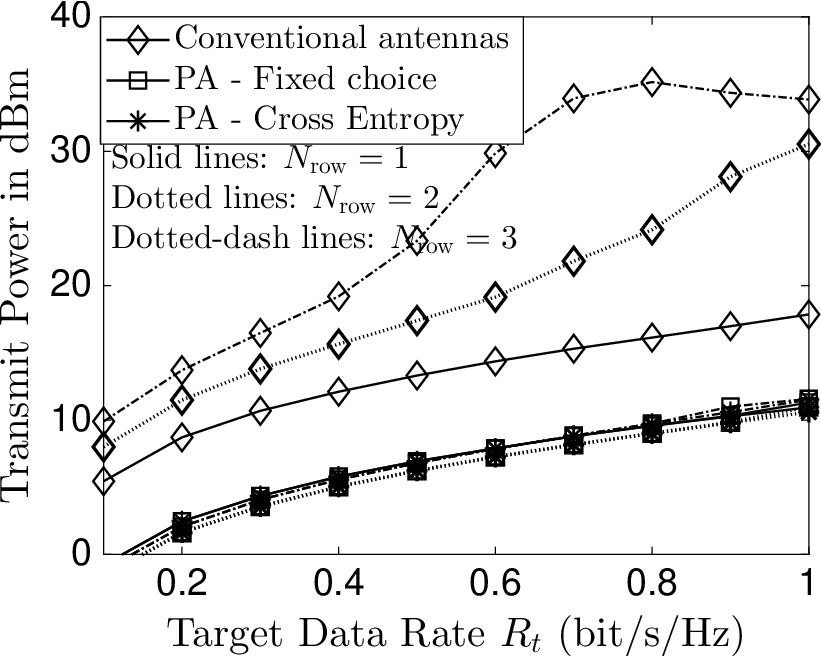}}   \vspace{-1em}
\end{center}
\caption{The overall transmit power consumption required by the considered multi-cell transmission schemes. The number of cell is $M=N_{\rm row}N_{\rm col}$, where $N_{\rm col}=2$.  \vspace{-1em} }\label{fig1}\vspace{-0.5em}
\end{figure}
  
In Fig. \ref{fig1}, the overall transmit power consumption required by the considered multi-cell transmission schemes is illustrated. Recall that there is an inevitable infeasibility situation for multi-cell transmission. Fig. \ref{fig1} shows two different approaches to mitigate the infeasibility situation. In particular, Fig. \ref{fig1a} is to use $P_{\rm max}=50$ dBm as the overall transmit power, if there is no feasible power allocation solution, and Fig. \ref{fig1b} is to simply discard those infeasible cases. Figs. \ref{fig1a} and \ref{fig1b} demonstrate that the use of pinching antennas can significantly reduce the overall transmit power consumption, regardless of which mitigation approach is used. The figures also show that the use of the fixed choice, $x_m^{\rm Pin}=x_m$, yields a performance close to the one achieved by the cross-entropy optimization algorithm, particularly if $R_t$ is small. In addition, Fig. \ref{fig1a} shows that as $R_t$ increases, the transmit power required by the conventional-antenna scheme quickly approaches $50$ dBm, which means that the benchmark scheme is prone to infeasibility under high data rate requirements. On the other hand,  there is only a small increase in transmit power for the pinching-antenna schemes, which shows the resilience of the pinching-antenna schemes to the infeasibility situation. Fig. \ref{fig1b} shows an interesting fact that for the case of $N_{\rm row}=3$, the transmit power required by the benchmark is reduced by increasing $R_t$ from $0.8$ bits/s/Hz to $1$ bits/s/Hz, which is due to the fact that Fig. \ref{fig1b} discards all the infeasible cases and the users' distances to their base stations might be very small for the preserved cases.

     \begin{figure}[!]\centering \vspace{-0.2em}
    \epsfig{file=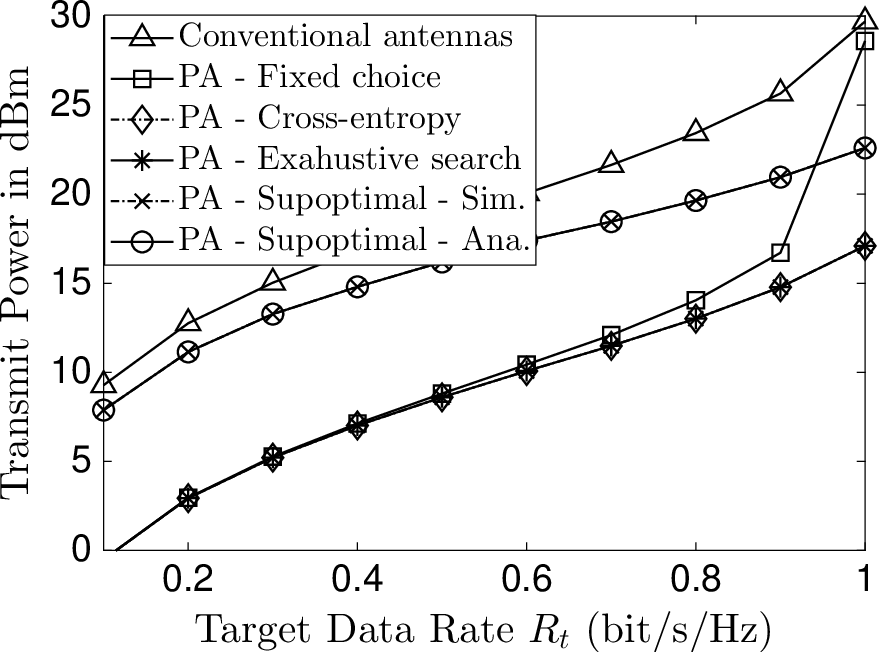, width=0.3\textwidth, clip=}\vspace{-0.5em}
\caption{The overall transmit power consumption achieved by the considered transmission schemes for the two-cell special case with clustered users, where $P_{\rm max}=50$ dBm. 
  \vspace{-1em}    }\label{fig2}   \vspace{-1em} 
\end{figure}

\begin{figure}[!]\centering \vspace{-2.5em}
    \epsfig{file=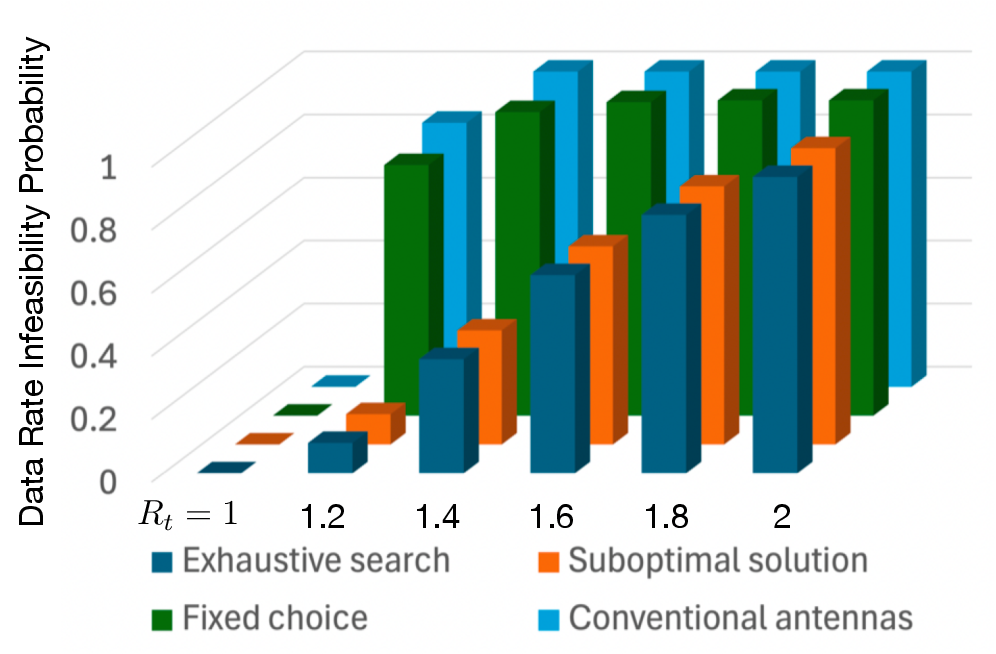, width=0.32\textwidth, clip=}\vspace{-0.5em}
\caption{The data rate infeasibility probability achieved by the considered transmission schemes for the two-cell special case with clustered users. 
  \vspace{-1em}    }\label{fig3}   \vspace{-0.5em} 
\end{figure}

Figs. \ref{fig2} and \ref{fig3} focus on the two-cell special case with clustered users, i.e., $x_1\sim \mathcal{U}(-2, 0)$ and $x_2\sim \mathcal{U}( 0,2)$, where $ \mathcal{U}(a, b)$ denotes a uniform distribution between $a$ and $b$. For the two-cell special case, an exhaustive search becomes feasible and hence is used to evaluate the optimality of the proposed cross-entropy optimization algorithm. As shown in Fig. \ref{fig2}, the proposed cross-entropy algorithm can realize exactly the same performance as the exhaustive search, which confirms the optimality of the proposed algorithm for the two-cell case. For the case with large $R_t$, the scheme of the fixed antenna location choice leads to a significant performance loss, which is due to the fact that for the fixed-choice scheme, the infeasibility condition becomes dominant for $R_t$ approaching $1$. Fig. \ref{fig2} also shows that the suboptimal solution proposed in Section \ref{subsection suboptimal} can always outperform the conventional-antenna benchmark, and it can also outperform the fixed-choice scheme for the cases with large $R_t$. Given the fact that the suboptimal solution can be obtained in a closed-form expression, it yields a better tradeoff between system performance and complexity, particularly when $R_t$ is large. The performance gain of the suboptimal solution over the fixed-choice scheme is mainly due to its resilience to the infeasibility situation, as confirmed by Fig. \ref{fig3}, where the infeasibility probability of the suboptimal solution is identical to that of the exhaustive search. 
\vspace{-0.5em}
\section{Conclusions}
In this letter, a practical multi-cell pinching-antenna scheme has been proposed, where cross-entropy optimization was employed for antenna location optimization. The simulation results revealed that interference management can be realized with low system overhead, and the use of multi-cell pinching-antenna systems leads to a quiet wireless world.  

\vspace{-0.5em}
\bibliographystyle{IEEEtran}
\bibliography{IEEEfull,trasfer}
  \end{document}